\documentclass[%
 reprint,
 amsmath,amssymb,
 aps,superscriptaddress,
]{revtex4-2}

\usepackage{graphicx}
\usepackage{dcolumn}
\usepackage{bm}

\usepackage{xcolor}
\usepackage{multirow}

\newcommand{\omegagw}{\Omega_{\rm{gw}}}

\newcommand{\mc}{\mathcal{M}_{c}}

\newcommand{\astro}{\theta_{a}}
\newcommand{\cosmo}{\theta_{c}}

\begin{document}

\preprint{APS/123-QED}


\title{Astrophysical and Cosmological Relevance \\ of the High-Frequency Features in the Stochastic Gravitational-Wave Background}

\author{Giulia Capurri}
\email{giulia.capurri@sissa.it}
\affiliation{SISSA, via Bonomea 265, I-34136, Trieste, Italy}
\affiliation{INFN-Trieste, Via Valerio 2, I-34127, Trieste, Italy}
\affiliation{IFPU, via Beirut 2, 34151, Trieste, Italy}

\author{Andrea Lapi}
\affiliation{SISSA, via Bonomea 265, I-34136, Trieste, Italy}
\affiliation{INFN-Trieste, Via Valerio 2, I-34127, Trieste, Italy}
\affiliation{IFPU, via Beirut 2, 34151, Trieste, Italy}
\affiliation{IRA-INAF, Via Gobetti 101, 40129 Bologna, Italy}

\author{Mario Spera}
\affiliation{SISSA, via Bonomea 265, I-34136, Trieste, Italy}
\affiliation{INFN-Trieste, Via Valerio 2, I-34127, Trieste, Italy}
\affiliation{INAF, Osservatorio Astronomico di Roma, Via Frascati 33, I-00040, Monteporzio Catone, Italy}

\author{Carlo Baccigalupi}
\affiliation{SISSA, via Bonomea 265, I-34136, Trieste, Italy}
\affiliation{INFN-Trieste, Via Valerio 2, I-34127, Trieste, Italy}
\affiliation{IFPU, via Beirut 2, 34151, Trieste, Italy}

\date{\today}

\begin{abstract}
The stochastic gravitational-wave background (SGWB) produced by merging neutron stars exhibits a peak in the kHz band. In this paper, we develop a theoretical framework to exploit this distinctive feature through a Markov Chain Monte Carlo analysis using a simulated dataset of SGWB measurements within this frequency range. The aim is to use the SGWB peak as an observable to constrain a set of astrophysical and cosmological parameters that accurately describe the sources of the SGWB. We examine how variations in these parameters impact the morphology of the SGWB and investigate the necessary sensitivity to effectively constrain them. Given our priors on astrophysical and cosmological parameters, and assuming a power-law integrated sensitivity curve of the order of $10^{-11}$ between 1 kHz and 5 kHz, we show that the values of the chirp mass and common envelope efficiency of the binary systems are retrieved with percent accuracy. Furthermore, the method allows for the reconstruction of the cosmological expansion history populated by these binaries, encompassing the Hubble constant, matter abundance, and the effective equation of state of dark energy. 
\end{abstract}


\maketitle

\section{\label{sec:intro}Introduction}
In the last years, the LIGO, Virgo, and KAGRA collaboration (LVK) has reported the detection of 90 gravitational-wave (GW) events from merging compact-object binaries, including binary black holes (BBH), binary neutron stars (BNS), and neutron star-black hole binaries (NSBH) \cite{2021arXiv211103606T, KAGRA:2021duu}. According to most astrophysical models, a few $10^5$ BBH mergers are expected to occur annually in the Universe, with BNS (and also NSBH) mergers being possibly even more frequent, reaching up to 10-100 times the BBH rate \cite{2019ApJ...881..157B,2021ApJ...907..110B,Santoliquido:2020axb,Santoliquido:2020bry,Santoliquido:2022kyu}. Hence, the detected events so far represent only a tiny fraction of the total.
The superposition of all the numerous unresolved events results in the stochastic gravitational-wave background (SGWB), which is a diffuse signal coming from all directions in the sky. If the sources responsible for the SGWB are extragalactic, the resulting signal shows a nearly homogeneous distribution with minimal anisotropies caused by the large-scale structure distribution of matter in the Universe \cite{Contaldi:2016koz,Cusin:2017fwz,Cusin:2018rsq,Cusin:2019jpv,Jenkins:2018nty,Jenkins:2018uac,Jenkins:2018kxc,Bertacca:2019fnt,Bellomo:2021mer,2021JCAP...11..032C}. 
Stochastic backgrounds are indistinguishable from instrumental noise in a single detector, but are correlated between pairs of detectors in ways that differ, in general, from instrumental noise \cite{Romano:2016dpx,Christensen:2018iqi,Renzini:2022alw}. As a consequence, extracting a SGWB signal requires cross-correlating the outputs of two or more detectors. Ultimately, a SGWB measurement can only be achieved using a network of multiple GW interferometers \cite{Thrane:2013oya, Alonso:2020rar}. 
The characterization of the SGWB usually relies on the energy density parameter $\omegagw(f)$ \cite{Allen:1997ad,Maggiore:1999vm,Mingarelli:2019mvk}: 

\begin{equation}
\omegagw(f ) 
= \frac{1}{\rho_{c}} \frac{d \rho_{\rm{gw}}(f )}{d \ln f },
\end{equation}
where $\rho_{\rm{gw}}$ is the SGWB energy density observed at the frequency $f$, and $\rho_{c} = 3H_{0}^{2}c^{2}/8\pi G$ is the critical energy density of the Universe \cite{Mingarelli:2019mvk}. While there has been no detection of the SGWB form ground-based interferometers yet, the LVK collaboration established the upper limit $\omegagw(f=25 \; \textrm{Hz}) \leq 3.4 \times 10^{-9}$ assuming a power-law SGWB with a spectral index of 2/3, which is the one expected for compact binary coalescences \citep{KAGRA:2021kbb}.
In June 2023, three major collaborations, the North American Nanohertz Observatory for Gravitational Waves (NANOGrav), the European Pulsar Timing Array (EPTA), and the Parkes Pulsar Timing Array (PPTA), jointly announced the first-ever detection of a SGWB through their pulsar timing array experiments \cite{NANOGrav:2023gor,EPTA:2023fyk,Reardon:2023gzh}. The origin of this SGWB remains uncertain, but one of the leading hypotheses points to the merging of supermassive BBHs at the centers of distant galaxies.
Indeed, there are numerous potential origins for the SGWB. These sources can be divided into two broad categories according to their nature: cosmological (e.g., cosmic inflation \cite{Bartolo:2016ami,Guzzetti:2016mkm,Caprini:2018mtu}, cosmological phase transitions \cite{Caprini:2007xq,Huber:2008hg,Caprini:2009yp,Hindmarsh:2013xza}, cosmic strings \cite{Damour:2004kw,Siemens:2006yp,Auclair:2019wcv}) and astrophysical (e.g., compact binary coalescences \cite{2011RAA....11..369R,  2011PhRvD..84h4004R, 2011PhRvD..84l4037M, 2011ApJ...739...86Z, 2013MNRAS.431..882Z, 2012PhRvD..85j4024W, 2016PhRvL.116m1102A, 2018PhRvL.120i1101A,Phinney:2001di,Regimbau:2008nj,Bavera:2021wmw,Perigois:2021ovr,Martinovic:2021fzj,Lehoucq:2023zlt}, rotating neutron stars \cite{1999MNRAS.303..258F,Marassi:2010wj}, exploding supernovae \cite{1999MNRAS.303..247F,Buonanno:2004tp,Crocker:2017agi}). 

\begin{figure}
\includegraphics[width=0.47\textwidth]{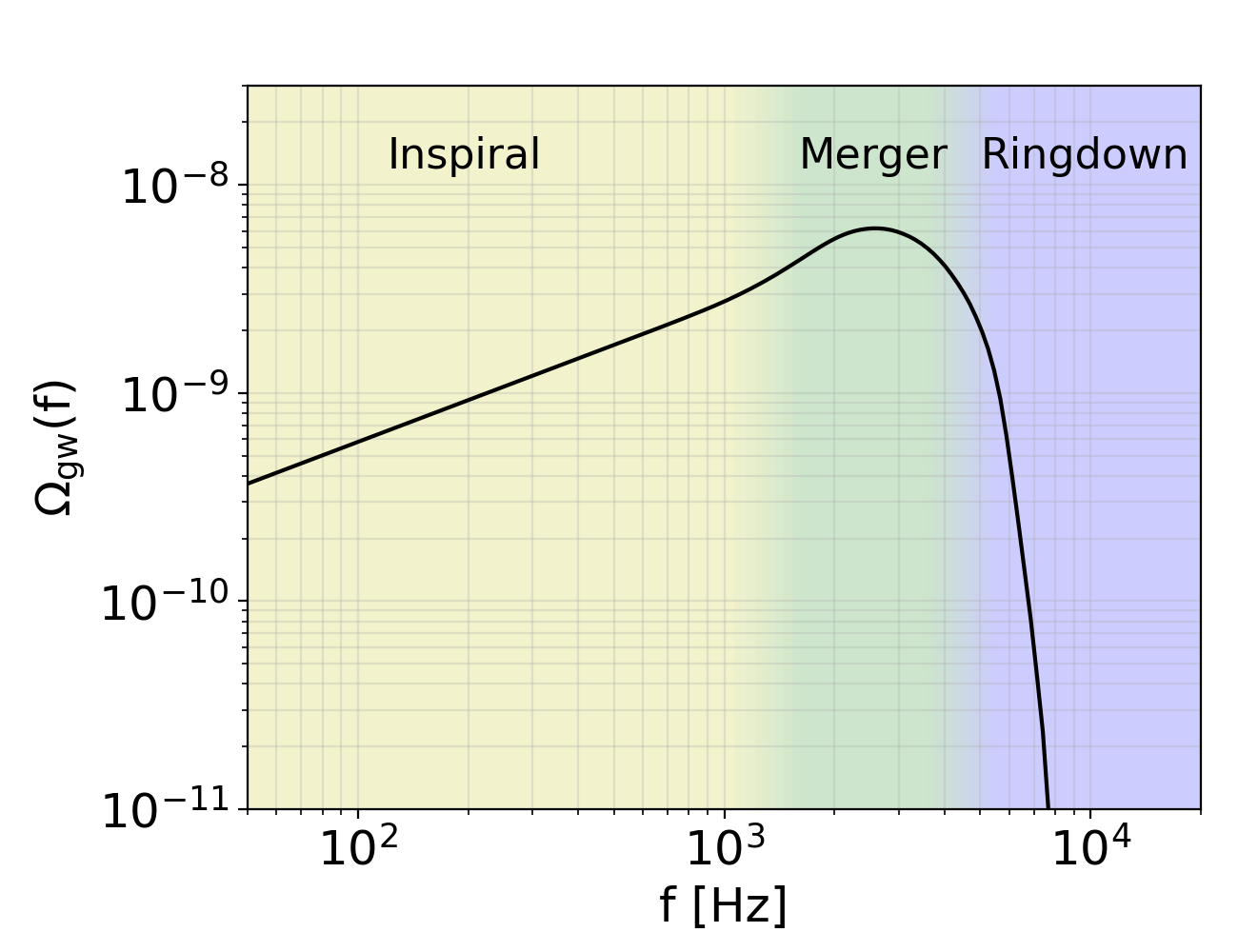}
\caption{\label{fig:phases} Typical frequency behavior of the energy density parameter as a function of the observed frequency. The power-law, peak and exponential cutoff regimes correspond to the inspiral, merger and ringdown phases, as explained in the text. This plot shows the SGWB produced by merging BNS, computed as in \cite{2021JCAP...11..032C,2022Univ....8..160C,2023ApJ...943...72C} according to the astrophysical prescriptions presented in \cite{2019ApJ...881..157B,2021ApJ...907..110B}. }
\end{figure}

The SGWB coming from compact binary coalescences is one of the main targets of present and forthcoming GW observatories. Indeed, such a signal (i) comes from all merging binaries since the beginning of stellar activity, and hence contains information about the entire population of sources; (ii) it is a tracer of the large-scale structure, as its anisotropies reflect those of the underlying dark matter distribution; (iii) it is dominant within the frequency band probed by ground-based interferometers \cite{KAGRA:2021kbb}. Consequently, an effective modeling of this specific component is needed to isolate other SGWB sources that might be present in the signal.
The amplitude and shape of the energy density $\omegagw(f)$ are primarily influenced by several astrophysical factors. Furthermore, another intriguing and largely unexplored characteristic of the SGWB from compact binary coalescences is its sensitivity to a set of cosmological parameters, including the Hubble parameter $H_{0}$. Therefore, relaying on a robust set of astrophysical and cosmological parameters is fundamental to provide an accurate description of the SGWB signal. The measurement of SGWB amplitude across multiple frequencies gives the opportunity to constrain the mentioned parameters. However, the majority of the binary coalescences building up the SGWB are in their inspiral phase between 10 Hz and a few hundred Hz. As a consequence, in this frequency regime, the energy density parameter follows a power-law behavior with a fixed slope, $\omegagw(f) \propto f^{2/3}$. Hence, in this regime, there is a strong degeneracy between the astrophysical and cosmological parameters that characterize the SGWB. Constraining the different parameters separately is even more difficult because their complex interplay shows up only as variations in the amplitude of the power-law.
In contrast, the scenario is different above a few hundred Hz. In this high-frequency regime, an increasingly larger portion of binaries evolves towards the merger and ringdown phases. Thus, the energy density parameter $\omegagw(f)$ shows a distinctive peak, as shown in Fig.~\ref{fig:phases}. The shape of the peak is influenced by a combination of astrophysical factors, such as the mass and redshift distribution of merging binaries, as well as cosmological factors, including the value of the Hubble parameter $H_{0}$, the matter content of the Universe, and the effective equation of state of dark energy. Therefore, the kHz range might contain additional information to better constrain the astrophysical and cosmological parameters describing the SGWB signal.

The aim of this paper is to study the information concealed in the peak of the SGWB. We will investigate how different sets of astrophysical and cosmological parameters affect the amplitude and shape of $\omegagw(f)$ in the high-frequency regime. Furthermore, we will show how a series of measurements within the kHz range can help to constrain these parameters. Finally, we will give some insights on the required sensitivity in the high-frequency regime needed to measure the $H_{0}$ parameter and possibly shed light on the Hubble tension \cite{DiValentino:2021izs}. There are alternative methods to estimate the value of the Hubble constant using GWs as, for example, with resolved signals. Each measurement provides the luminosity distance to the source, while the corresponding redshift can be obtained using various approaches, including the redshifted masses and a galaxy catalog \cite{LIGOScientific:2021aug, Maggiore:2019uih, Branchesi:2023mws}. The value of $H_0$ is then inferred from the $d_L-z$ relation. The method presented in this paper represents a completely independent approach.

This paper is organized as follows. In Section \ref{sec:dependencies}, we briefly recall the derivation of the SGWB energy density parameter for binary coalescences. Then, we identify a set of astrophysical and cosmological parameters suitable for describing a specific family of coalescing binaries (i.e., BNSs) and study how different values of such parameters affect $\omegagw (f)$. In Section \ref{sec:methods}, we describe our methodology for exploiting the high-frequency features of the SGWB through a Markov Chain Monte Carlo analysis using a simulated data set of SGWB measurements. In Section \ref{sec:results}, we study how well different input values of the astrophysical and cosmological parameters are retrieved with the MCMC analysis. Finally, we discuss our findings and draw our conclusions in Section \ref{sec:discussion}.

\section{\label{sec:dependencies}Study of physical dependencies}

\begin{figure*}
\includegraphics[width=\textwidth]{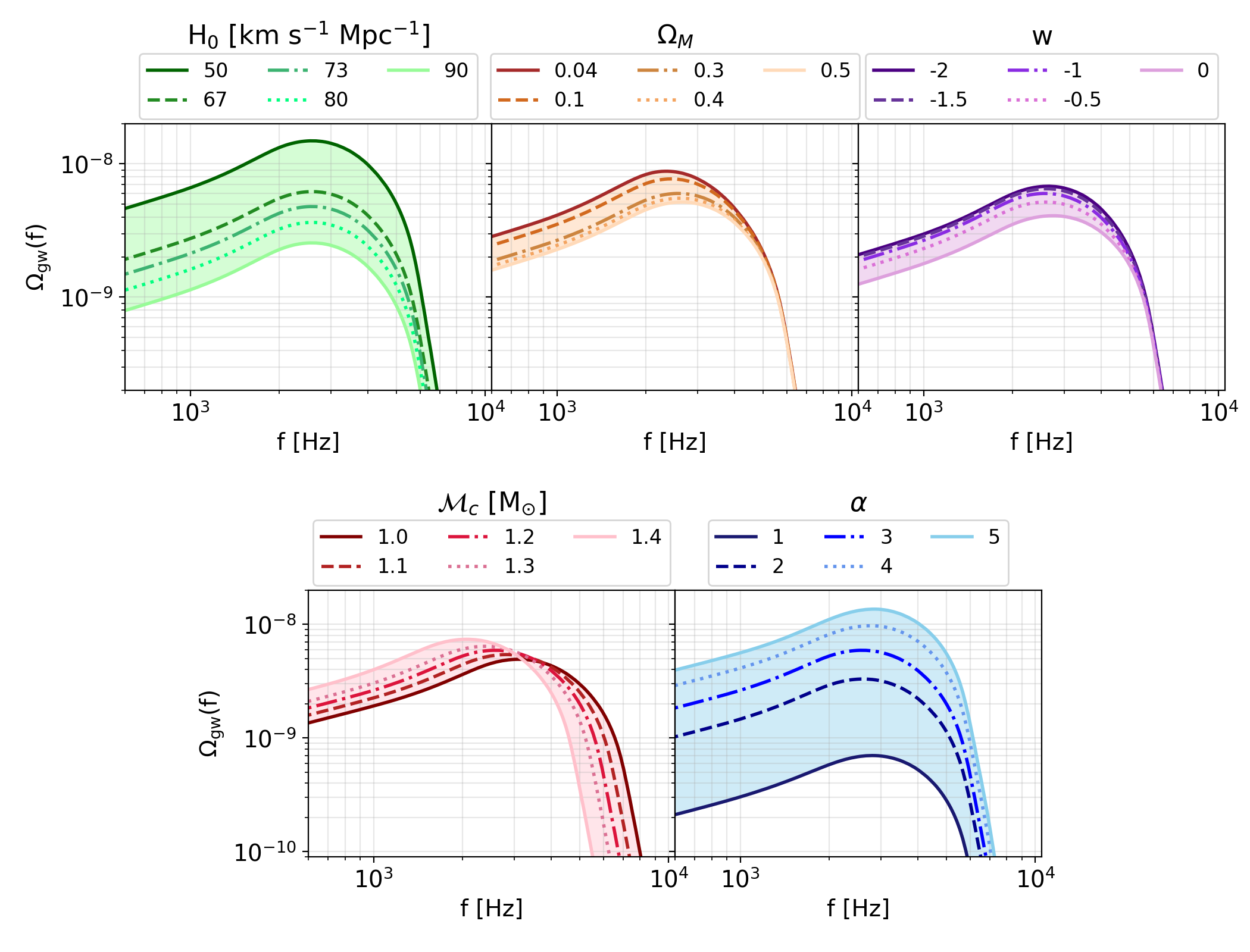}
\caption{\label{fig:params} The energy density $\omegagw(f)$ of the SGWB produced by merging BNS systems exhibits different behaviors for various values of the cosmological and astrophysical parameters considered in our analysis: $\cosmo = \{H_{0}, \Omega_{M}, w\}$ and $\theta_{a} = \{\mc, \alpha \}$. In each panel, we explore the effect of varying a single parameter while keeping the others fixed at their reference values: $H_{0}^{\rm{ref}} = 68$ $\rm{km} \, \rm{s}^{-1} \rm{Mpc}^{-1}$, $\Omega_{M}^{\rm{ref}} = 0.315$, $w^{\rm{ref}}=-1$, $\mc^{\rm{ref}} = 1.2$ M$_{\odot}$, and $\alpha^{\rm{ref}} = 3$. }
\end{figure*}

In this section, we give an overview of the astrophysical and cosmological dependencies of the SGWB energy density parameter. 
Following Refs. \cite{2011RAA....11..369R,  2011PhRvD..84h4004R, 2011PhRvD..84l4037M, 2011ApJ...739...86Z, 2013MNRAS.431..882Z, 2012PhRvD..85j4024W, 2016PhRvL.116m1102A, 2018PhRvL.120i1101A,Phinney:2001di,Regimbau:2008nj}, $\omegagw(f)$ can be re-written as:

\begin{equation} \label{eq:omega_gw_def}
    \omegagw(f) \equiv \frac{1}{\rho_{c}} \frac{d \rho_{\rm{gw}}(f )}{d \ln f }
    = \frac{f}{\rho_{c}} \frac{d^{2} \mathcal{E}_{\rm{gw}}}{dVdf} 
    = \frac{f}{\rho_{c} c} \frac{d^{3} \mathcal{E}_{\rm{gw}}}{dSdtdf}, 
\end{equation}
where $\mathcal{E}_{\rm{gw}}$ is the total energy carried by the stochastic background, so that $d^{3} \mathcal{E}_{\rm{gw}} / dSdtdf $ is the total energy flux per unit time and frequency in the observer frame. By expanding Eq. (\ref{eq:omega_gw_def}), we get

\begin{equation} \label{eq:flux_rate}
    \omegagw(f) = \frac{f}{\rho_{c} c} \int dz d\astro \, p(\astro) \, F(f, z| \astro) \, \frac{d \dot{N}}{dz}(z|\astro). 
\end{equation}
In Eq. (\ref{eq:flux_rate}), $p(\astro)$ is the probability distribution of the source astrophysical parameters, $\astro$. $F(f, z| \astro)$ is the averaged energy flux per unit observed frequency emitted by coalescing binaries located at redshift $z$ and characterized by the astrophysical parameters $\astro$: 

\begin{equation}
  F(f, z| \astro)  = \frac{ \dfrac{dE_{\rm{gw}}}{df} (f| \astro)}{4\pi d_{L}^{2}(z|\cosmo)} = \frac{ \dfrac{dE_{\rm{gw}}}{df_{s}} (f_{s}| \astro)}{4\pi r^{2}(z|\cosmo) (1+z)},  
\end{equation}
where $dE_{\rm{gw}}/df$ is the emitted gravitational spectral energy and $f_{s} = f (1+z)$ is the frequency in the source frame. $d_{L}(z| \cosmo)$ and $r(z| \cosmo)$ are the luminosity distance and the proper distance, respectively, and depend on the adopted cosmology, defined by the cosmological parameters $\cosmo$. The last term of the integral in Eq.~(\ref{eq:flux_rate}) is the rate of mergers per redshift interval. This quantity can be expressed in terms of the intrinsic merger rate per unit comoving volume, $R(z|\astro)$, as follows: 

\begin{equation}
   \frac{d \dot{N}}{dz}(z|\astro) =  R(z|\astro) \frac{dV}{dz}, 
\end{equation}
with 

\begin{equation}
    \frac{dV}{dz} = \frac{4\pi c \, r^{2}(z| \cosmo)}{H(z| \cosmo)}, 
\end{equation}
where $H(z| \cosmo) = H_{0} h(z| \cosmo)$ is the Hubble rate.
By combining everything together, we obtain the well-known expression for the SGWB energy density parameter, as reported in Refs. \cite{2011RAA....11..369R,  2011PhRvD..84h4004R, 2011PhRvD..84l4037M, 2011ApJ...739...86Z, 2013MNRAS.431..882Z, 2012PhRvD..85j4024W, 2016PhRvL.116m1102A, 2018PhRvL.120i1101A,Phinney:2001di,Regimbau:2008nj}:
\begin{widetext}
\begin{equation} \label{eq:omega_gw}
    \omegagw(f) = \frac{8\pi G f}{3 H_{0}^{3} c^{2}} \int dzd\astro \, p(\astro) \, \frac{dE_{\rm{gw}}}{df_{s}} (f_{s},z| \astro) \frac{R(z|\astro)}{(1+z)h(z|\cosmo)}.
\end{equation}
\end{widetext}

\begin{figure*} 
\includegraphics[width=\textwidth]{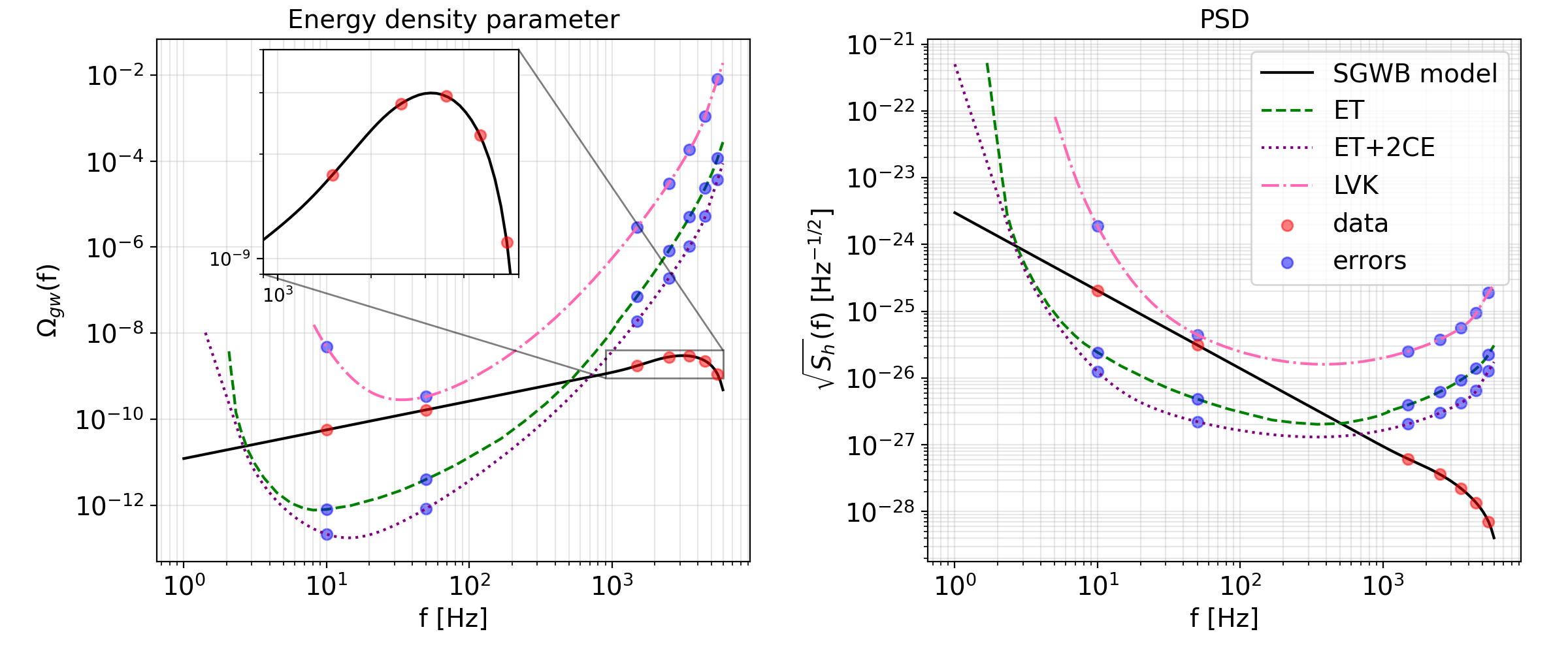}
\caption{\label{fig:et_data_errors} The solid line in the left(right) panel represents our model of $\omegagw(f)$ ($S_{h}^{1/2}(f)$) for the SGWB produced by coalescing BNSs. As explained in the text, the model is based on empirical galactic prescriptions (B18 FMR model from \cite{Santoliquido:2022kyu}) and two set of parameters, $\astro$ and $\cosmo$. The coloured lines are the power-law integrated sensitivity curves (PLS) of LVK, Einstein Telescope (ET), and ET in combination with two Cosmic Explorers, one in the US and one in Australia. The red and blue points represent a hypothetical data-set with the corresponding errors.}
\end{figure*}

The adopted cosmological model affects $\omegagw(f)$ through the $H_{0}$ parameter. Furthermore, the cosmology influences the behavior of $h(z| \cosmo)$, which has distinct functional forms depending on the adopted cosmological scenario. In this study, we use a standard flat cosmology, with the Hubble parameter $H_{0}$, the matter density parameter $\Omega_{M}$, and the dark energy equation of state $w$ as free parameters. In this scenario, the expression for $h(z| \cosmo)$, with $\cosmo = \{H_{0}, \Omega_{M}, w\}$, is: 

\begin{equation}
    h(z| \cosmo) = \sqrt{\Omega_{M}(1+z)^{3} + \Omega_{\Lambda}(1+z)^{3(1+w)}},
\end{equation}
where $\Omega_{\Lambda} = 1-\Omega_{M}$ for the flatness requirement. The top panels of Fig.~\ref{fig:params} show the dependence of $\omegagw(f)$ on the set of cosmological parameters. From Fig. \ref{fig:params}, it is apparent that the SGWB energy density is most sensitive to $H_{0}$, as $\omegagw \propto H_{0}^{-3}$. This means that higher values of $H_{0}$ result in a reduced SGWB amplitude because a faster cosmic expansion leads to a more significant dilution of the energy density. 
However, assessing the sensitivity of a SGWB measurement to the Hubble parameter using $\omegagw(f)$ may not be the most suitable approach. Notably, a substantial dependence on $H_{0}$ arises from the presence of $\rho_{c}$ in the definition of $\omegagw(f)$. As a consequence, instead of relying on $\omegagw (f)$, we will use the spectral density $S_{h}(f)$, which is directly measured by GW detectors. Indeed, a detector produces an output of the measured GW strain, $h(t)$. From the correlation of the outputs of two detectors one can measure the root mean square of the strain, $h^{2}_{\rm{rms}}$, or, equivalently, the power spectral density (PSD) $S_{h}(f)$, which is defined through (see e.g. \cite{Moore:2014lga,Romano:2016dpx}): 

\begin{equation}
    h^{2}_{\rm{rms}} = \Bigl \langle \sum_{ij} h_{ij} h_{ij} \Bigl \rangle = \int_{0}^{\infty} df S_{h}(f).
\end{equation}
The PSD and the energy density parameter are related through 

\begin{equation} \label{eq:psd}
     S_{h}(f) =  \frac{3 H_{0}^{2} }{2 \pi^2 f^{3}}  \omegagw(f), 
\end{equation}
so that 

\begin{widetext}
\begin{equation} \label{eq:psd_long}
    S_{h}(f) = \frac{4G}{\pi H_{0} c^{2}} f^{-2} \int dzd\astro \, p(\astro) \, \frac{dE_{\rm{gw}}}{df_{s}} (f_{s},z| \astro) \frac{R(z|\astro)}{(1+z)h(z|\cosmo)}.
\end{equation}
\end{widetext}

The astrophysical dependencies of $\omegagw(f)$ are embedded within the merger rate and the GW spectral energy, and are more complex than the cosmological ones. Indeed, the intricate interplay of numerous processes, spanning from the physics of stars and binary systems to that of the host galaxies, decisively shapes the formation, evolution and merger of compact binaries. As a consequence, capturing all the involved processes with a limited set of parameters is challenging. 
The difficulty is particularly pronounced in the case of BBHs, as their mass and redshift distributions show complex and distinctive features, heavily influenced by a multitude of astrophysical factors, including the evolution of massive stars (e.g. pair instability, core collapse, natal kicks, etc.), different binary formation channels (e.g. isolated, dynamical, etc.), binary evolution processes (e.g. stable mass transfer, common envelope, etc.), and the metallicity and star formation rate of the galactic environment (see, e.g, Refs. \cite{2019ApJ...881..157B,2021ApJ...907..110B,Santoliquido:2020bry,Santoliquido:2020axb,Santoliquido:2022kyu,Sicilia:2021gtu} for a comprehensive description of relevant physics at play).
For BNSs, the complexity level is significantly reduced. Firstly, uncertainties concerning the galactic environment are smaller compared to the BBH case. Indeed, BNS systems are minimally affected by metallicity variations, with their evolution mainly depending on the galaxy main sequence (i.e the relation between stellar mass and star formation rate), which is empirically well-constrained \cite{Speagle:2014loa,2023MNRAS.519.1526P}. Secondly, while there are larger uncertainties on the stellar side, the mass spectrum of neutron stars sharply peaks around $1.3 \, M_{\odot}$, especially when considering the extensive sample of electromagnetic observations of galactic neutron stars \cite{Ozel:2012ax,Farrow:2019xnc}. However, GW detections of neutron stars appear to be compatible with a broader mass distribution, possibly extending to higher masses \cite{Landry:2021hvl}. Notably, the six neutron stars observed through GW events so far seem to be generally heavier than those observed through electromagnetic radiation. For instance, GW190425 likely involves a neutron star of approximately $2 \, M_{\odot}$ \cite{LIGOScientific:2020aai}. Nevertheless, given the limited number of GW detections, we find it more practical for our analysis to rely on the mass distribution established through electromagnetic observations.
For these reasons, we have decided to work with the SGWB produced by BNS systems, which can be described with a reasonable number of astrophysical parameters. Specifically, we focus on the stellar domain and characterize the BNS population through two key astrophysical parameters, $\theta_{a} = \{\mc, \alpha \}$, where $\mc$ represents the value at which the chirp mass distribution peaks, and $\alpha$ denotes the common envelope efficiency parameter \cite{Giacobbo:2018etu,Fragos:2019box}. The common envelope phase occurs when the envelope of a giant star engulfs the companion. This phase is crucial for the evolution of the binary system and the subsequent GW emission, as the friction between the companion star and the giant’s envelope transforms orbital energy and angular momentum into heat and angular momentum of the common envelope, shrinking the binary orbit. The efficiency of this transfer is encoded in the value of $\alpha$, which, along with the chirp mass, constitutes the parameter that predominantly determines the merger rates of BNSs \cite{Santoliquido:2020axb}. There are other physical mechanisms, described by additional parameters, that affects the merger rates. However, either their impact is smaller in comparison to varying $\alpha$ and $\mathcal{M}_c$, or the range of values they can assume to maintain compatibility with the LVK merger estimates is narrow (see, e.g., Figg. 3-5 of Ref. \cite{Santoliquido:2020axb}). Therefore, for many purposes is reasonable to use only $\alpha$ and $\mathcal{M}_c$ as free parameters and assign fixed fiducial values to the other ones, ensuring the retrieval of the local LVK merger rates. This approach is also employed in Ref. \cite{Santoliquido:2022kyu}, from which we adopt the model for the merger rates. Notably, these considerations specifically apply to BNSs. For BBHs, the significant degeneracy among the parameters allow them to vary within broad ranges of values while still yielding merger rate predictions consistent with the local LVK estimates.
On the galactic side, instead, given the increasing confidence in forthcoming observational constraints, we rely on empirical, data-driven prescriptions, based on multi-band measurements of the galaxy main sequence and metallicity. In particular, we establish a fixed fiducial scenario for metallicity and main sequence, the B18 FMR model in Ref. \cite{Santoliquido:2022kyu}.
In the lower panel of Fig. \ref{fig:params}, we show how $\omegagw(f)$ depends on the parameters $\theta_{a}$. Notably, different values of $\mc$ cause a shift in the peak's position, as BNS populations with different masses merge at different typical frequencies. Conversely, varying $\alpha$ leads to a significant change in the amplitude of the SGWB, as it directly affects the number of merging BNS binaries.

\section{\label{sec:methods}Methods}

\begin{figure}
\includegraphics[width=0.5\textwidth]{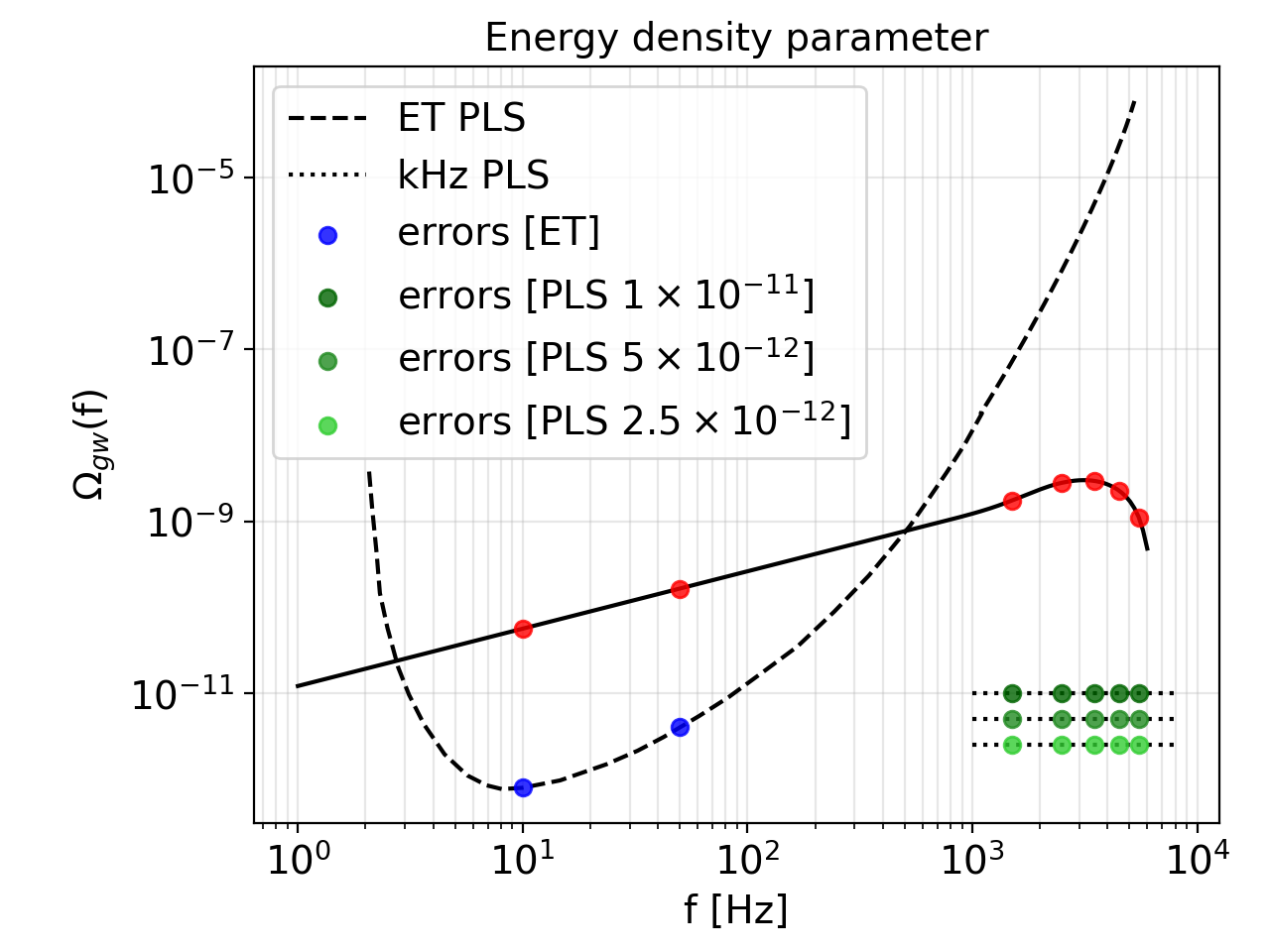}
\caption{\label{fig:kHz_data_errors} Example of the mock data-set that we use in our analysis. $\omegagw(f)$ is computed using a specific set of input values for our parameters $\astro$ and $\cosmo$ (see Table~\ref{tab:fiducial_prior}). The data points, in red, are obtained sampling $\omegagw(f)$ at seven distinct frequencies. The associated errors are fixed to the ET 1$\sigma$-PLS in the 1-100 Hz regime (blue points), and at progressively low arbitrary values in the kHz regime (green points).}
\end{figure}

In this work, we characterize the SGWB using its PSD, $S_{h}(f)$, which is linked to the energy density parameter, $\omegagw(f)$, through Eq.~(\ref{eq:psd}). As already mentioned, we adopted this approach for two reasons: \textit{i)} $\omegagw(f)$ introduces a further dependence on $H_{0}$, potentially affecting the relationship between the SGWB amplitude and the Hubble parameter, and \textit{ii)} the PSD is more directly related to the GW strain, the quantity measured by detectors. However, since the results about the SGWB are usually expressed in terms of $\omegagw(f)$, we also present our results in terms of $\omegagw(f)$ instead of $S_{h}(f)$.
As a preliminary step, we calculate the PSD at different frequencies in the range $[10 \, \textrm{Hz}-5.5 \,\textrm{kHz}]$, considering different sets of astrophysical and cosmological parameters. The PSD values are our mock measurements, and we also associate an error to each of them. The error is calculated by computing the $1\sigma$ power-law integrated sensitivity curve (PLS) \cite{Thrane:2013oya} for a specific network of detectors, assuming an observation time $T = 1$ yr. 
The value of the $1\sigma$-PLS at each frequency represents the amplitude of a power-law SGWB with a signal-to-noise ratio of 1, providing a reasonable estimation of the error for our measurements. 
Once we have our data-set with errors, we perform a Markov Chain Monte Carlo (MCMC) analysis to retrieve the input values of the astrophysical and cosmological parameters that we used to generate the data. We use the code \texttt{emcee}, which is an MIT licensed pure-Python implementation of Goodman $\&$ Weare’s Affine Invariant MCMC Ensemble sampler \cite{Foreman_Mackey_2013}. 

In Fig.~\ref{fig:et_data_errors}, we show a collection of mock data points along with their corresponding errors, computed in the context of a detection with three different networks: (i) the current network of second-generation instruments, LIGO, Virgo, and KAGRA (LVK) \cite{2015CQGra..32g4001L,2015CQGra..32b4001A,2012CQGra..29l4007S} at design sensitivity (post-O5), (ii) the third-generation detector Einstein Telescope (ET) \cite{Branchesi:2023mws,Maggiore:2019uih}, and (iii) an extended network composed of ET and two Cosmic Explorer (CE) detectors \cite{Reitze:2019iox,Evans:2021gyd}, one in the US and one in Australia. The left panel of Fig.~\ref{fig:et_data_errors} shows the expected $\omegagw(f)$ based on our fiducial values of the astrophysical and cosmological parameters, as reported in Table~\ref{tab:fiducial_prior}. 
\begin{table}[b]
\caption{\label{tab:fiducial_prior}%
Fiducial values and prior intervals for our astrophysical and cosmological parameters. All the priors are flat, except from the one for $\mc$, which is assumed to be a Gaussian with $\sigma_{\mc} = 0.2$ centered around $\bar{\mathcal{M}}_{c} = 1.2$.}
\begin{ruledtabular}
\begin{tabular}{cccc}
\textrm{Parameter}&
\textrm{Fiducial value(s)}&
\textrm{Prior interval}&
\textrm{Units}\\
\colrule
$\mc$ & 1.25 & [1,1.5] & $M_{\odot}$\\ [3pt]
$\alpha$ & 3.8 & [1,5] & / \\ [3pt]
\multirow{ 2}{*}{$H_{0}$} & 67.4 & \multirow{ 2}{*}{[50,90]} & \multirow{ 2}{*}{km s$^{-1}$Mpc$^{-1}$} \\ 
 & 73 &  & \\ [3pt]
$\Omega_{M}$ & 0.315 & [0.04, 0.5] & / \\ [3pt]
$w$ & -1.5 & [-2,0] & / \\
\end{tabular}
\end{ruledtabular}
\end{table}
We also show our mock data (red points), which are given by the expected values of $\omegagw(f)$ at specific frequencies where measurements are assumed to be taken at: $f = 10$ Hz, 50 Hz, $1.5 \times 10^{3}$ Hz, $2.5 \times 10^{3}$ Hz, $3.5 \times 10^{3}$ Hz, $4.5 \times 10^{3}$ Hz, and $5.5 \times 10^{3}$ Hz. The first two frequencies are strategically chosen within the region where LVK, ET, and CE have maximum sensitivity to stochastic backgrounds. These data points are crucial for constraining the amplitude of the SGWB. Instead, the five data points in kHz range are essential to characterize the peak of the SGWB, as it is shown in the zoomed-in region in the left plot of Fig.~\ref{fig:et_data_errors}. The errors of our mock measurements (blue points) match the values of the PLS of the considered detector network at the observed frequencies. In the right panel of Fig.~\ref{fig:et_data_errors}, we present the same quantities as in the left panel, but expressed in terms of the PSD using Eq.~(\ref{eq:psd}).
From Fig.~\ref{fig:et_data_errors}, it is apparent that LVK at design sensitivity will only marginally detect the SGWB. In contrast, the improved sensitivity of ET will allow an even better characterization of the SGWB in the frequency range from a few Hz to a few hundred Hz. Moreover, the PLS shown in the plot refers to the current expectations for ET sensitivity. Once online, the detector will undergo continuous upgrades, similar to LVK, that will improve the detector's sensitivity. Combining ET with other third-generation detectors, such as CE, will further enhance the overall sensitivity, even though the characterization of the high-frequency peak of the SGWB from BNSs is unlikely for all ground-based interferometers.

\begin{figure*}
\includegraphics[width=0.96\textwidth]{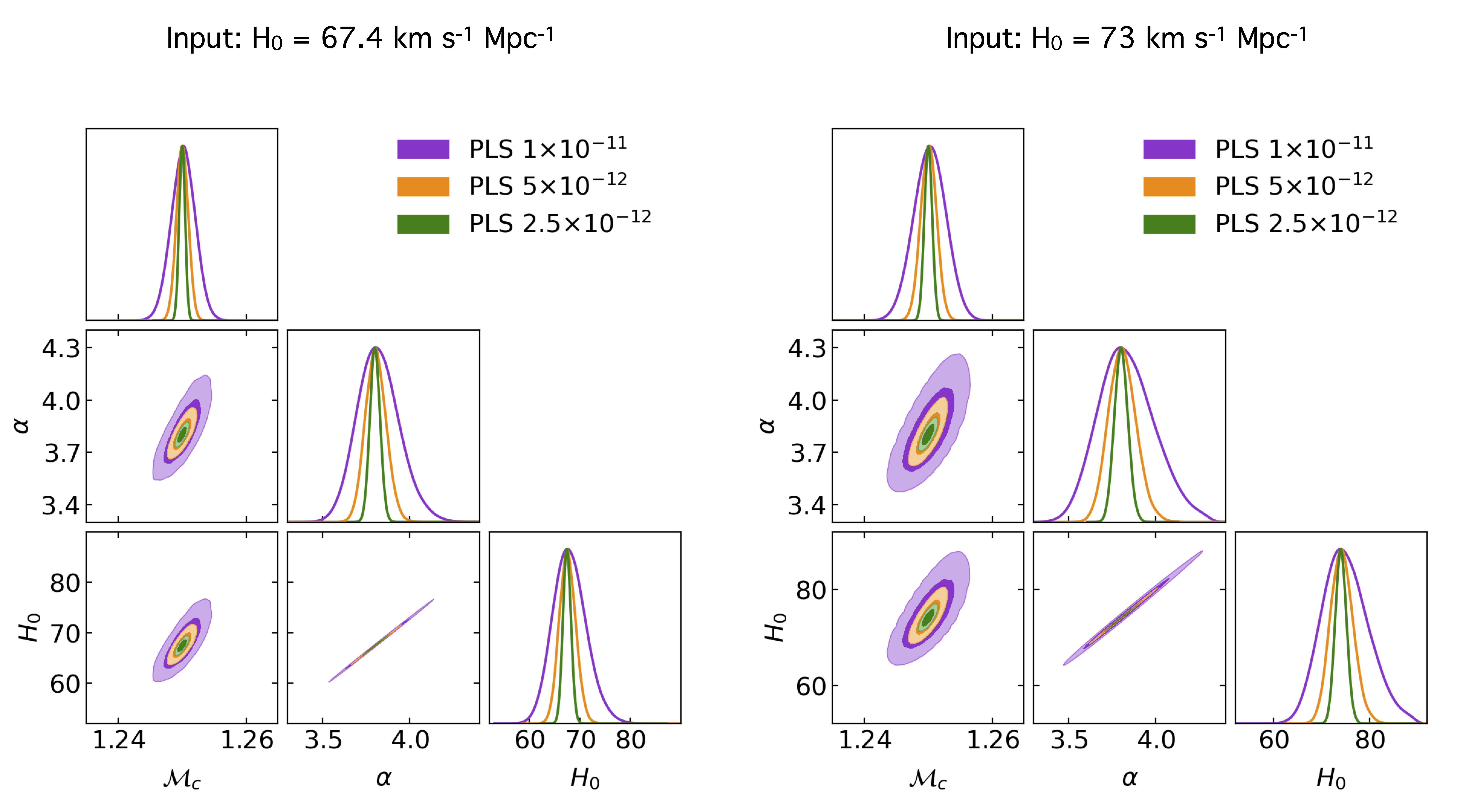}
\caption{\label{fig:posteriors_h0} Joint constraints (68$\%$ and 95$\%$ confidence regions) and marginalized posterior distributions on $\mc$, $\alpha$ and $H_{0}$, for two sets of input values $\{ 1.25 \, M_{\odot}, 3.8, 67.4\, \rm{km} \, \rm{s}^{-1} \rm{Mpc}^{-1}  \}$ and $\{ 1.25 \, M_{\odot}, 3.8, 73\, \rm{km} \, \rm{s}^{-1} \rm{Mpc}^{-1}  \}$, and three different sensitivity levels in the kHz range.} 
\end{figure*}

The primary objective of this paper is to build up a science case to assess whether the SGWB measured in the kHz band can serve as a reliable observable to constrain cosmological and astrophysical parameters. To accomplish this goal, we manually fix the sensitivity in the kHz band so to reach a good level of constraining power on the astrophysical and cosmological parameters. Fig.~\ref{fig:kHz_data_errors} shows a typical data-set that we used for our theoretical analysis, with manually fixed errors in the kHz band (green points). For the first two data points, we use the errors associated with ET.
We generate different data-sets using the fiducial values of the parameters $\cosmo$ and $\astro$, as reported in Table~\ref{tab:fiducial_prior}. The real values $\mc, \alpha$ and $w$ are highly uncertain, thus we randomly pick their fiducial values inside the prior ranges typically used in the literature \cite{Giacobbo:2018etu,Fragos:2019box,Giacobbo:2018hze,Laghi:2021pqk,Planck:2013pxb,Planck:2015fie}. In contrast, for $H_{0}$ and $\Omega_{M}$ we take the latest values obtained by Planck \cite{2020A&A...641A...6P}. For $H_{0}$, we also consider an additional fiducial value, corresponding to the local measurement from Cepheid variables and Type Ia supernovae \cite{Riess:2021jrx}. All the priors distributions are flat, except that of $\mc$, which is a Gaussian with $\sigma_{\mc} = 0.2\,M_{\odot}$ centered around $1.2\,M_{\odot}$. We then perform an MCMC to retrieve the input values of our parameters. Finally, we study the amplitude of the posterior contours for different choices of the kHz PLS, which give an estimate of the constraining power of our observable.

Finally, we emphasize that following this preliminary science case, we plan to apply our methodology to more advanced scenarios. For example, the description of BNS systems could be enhanced by including the dependence on the neutron star equation of state. The equation of state affects the masses of the binary components and the GW waveforms, both of which contribute to the SGWB energy density.
We also plan to extend our study to BBHs, which are expected to produce a SGWB with a peak at lower frequencies (a few hundred Hz). As mentioned earlier, the BBH case requires a larger number of parameters to be described because the properties and evolution of such systems heavily depend on the metallicity and various formation channels. Furthermore, since the next-generation GW detectors will resolve the majority of coalescing BBHs, implementing our methodology for such systems will involve considering the residual SGWB, obtained by excluding all resolved events from the energy density computation.

\section{\label{sec:results}Results}

\begin{figure}
\includegraphics[width=0.48\textwidth]{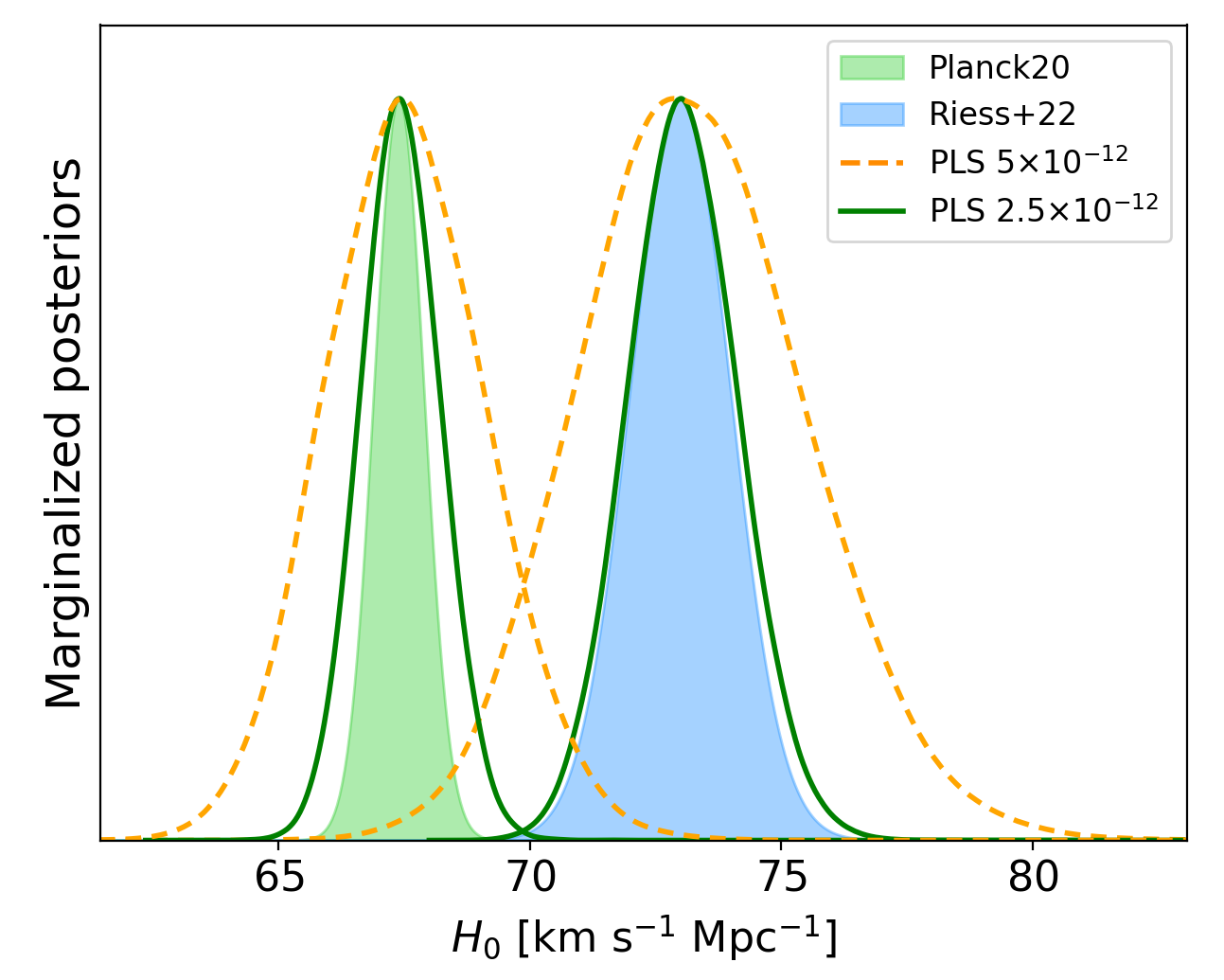}
\caption{\label{fig:h0_comparison} Comparison of various posterior constraints on $H_{0}$. The green and blue shaded regions represent the constraints from Planck data \cite{2020A&A...641A...6P} and local Cepheids and Type Ia supernovae \cite{Riess:2021jrx}, respectively. The orange and green lines are the constraints obtained using the “reduced” parameter space ${M_c, \alpha, H_0}$ at two different kHz sensitivity levels.}
\end{figure}

In Fig.~\ref{fig:posteriors_h0}, we show the joint constraints (68$\%$ and 95$\%$ confidence regions) and marginalized posterior distributions on $\mc$, $\alpha$ and $H_{0}$ for two sets of input values, $\{ 1.25 \, M_{\odot}, 3.8, 67.4\, \rm{km} \, \rm{s}^{-1} \rm{Mpc}^{-1}  \}$ and $\{ 1.25 \, M_{\odot}, 3.8, 73\, \rm{km} \, \rm{s}^{-1} \rm{Mpc}^{-1}  \}$. In Table~\ref{tab:perc_constraints_3}, we report the associated marginalized percentage constraints at the 68$\%$ confidence level.
\begin{table}[b]
\caption{\label{tab:perc_constraints_3}%
Marginalized percentage constraints at the 68$\%$ confidence level on $\mc$, $\alpha$ and $H_{0}$, with input values $\{ 1.25 \, M_{\odot}, 3.8, 73\, \rm{km} \, \rm{s}^{-1} \rm{Mpc}^{-1}  \}$, for the three different kHz sensitivity levels.}
\begin{ruledtabular}
\begin{tabular}{cccc}
 & PLS $1\times 10^{-11}$ & PLS $5\times 10^{-12}$ & PLS $2.5\times 10^{-12}$\\
\colrule
$\mc$    &   $0.2\%$  &  $0.1\%$   &   $0.1\%$  \\ [3pt]
$\alpha$ &  $4.2\%$   &   $2.1\%$  &  $1.0\%$   \\ [3pt]
$H_{0}$  &  $6.5\%$   &  $3.2\%$   &  $1.6\%$   \\ [3pt]
\end{tabular}
\end{ruledtabular}
\end{table}
The input values for $H_{0}$ are chosen to match the most recent Planck \cite{2020A&A...641A...6P} and local \cite{Riess:2021jrx} estimates, respectively. For both sets of input parameters, we explore the constraining power of our mock data-set for different kHz sensitivities. We find that a PLS $ = 1\times 10^{-11}$ is the poorest sensitivity for which the data have some constraining power on the three considered parameters. For higher values of the PLS, the posteriors are dominated by the priors and thus become uninformative. At PLS $ = 1\times 10^{-11}$, instead, the astrophysical parameters are retrieved quite well, so as the Hubble parameter. At this sensitivity level, however, it is not possible to distinguish between the two $H_{0}$ input values with enough significance. The determination of $H_{0}$ is further complicated by the significant degeneracy with $\alpha$, given that both parameters impact the amplitude of the SGWB, while leaving its shape mostly unvaried (refer to Fig.~\ref{fig:params}). Nevertheless, any degeneracy among the parameters is comprehensively accounted for within our Bayesian approach, and is reflected in the dispersion of the associated posterior constraints.
As expected, the constraining power increases for lower values of the PLS. In particular, with a PLS = $5\times 10^{-12} \,(2.5\times 10^{-12})$ it is possible to distinguish the two conflicting values of the Hubble parameter at $1(2)\sigma$. As shown in Fig. \ref{fig:h0_comparison}, the posterior constraints on $H_{0}$ obtained with the lowest PLS have widths comparable to existing constraints from Planck data \cite{2020A&A...641A...6P} and local Cepheids and Type Ia supernovae \cite{Riess:2021jrx}. Therefore, an $H_{0}$ estimate of such precision could potentially rule out one of the two existing constraints.

\begin{figure}
\includegraphics[width=.5\textwidth]{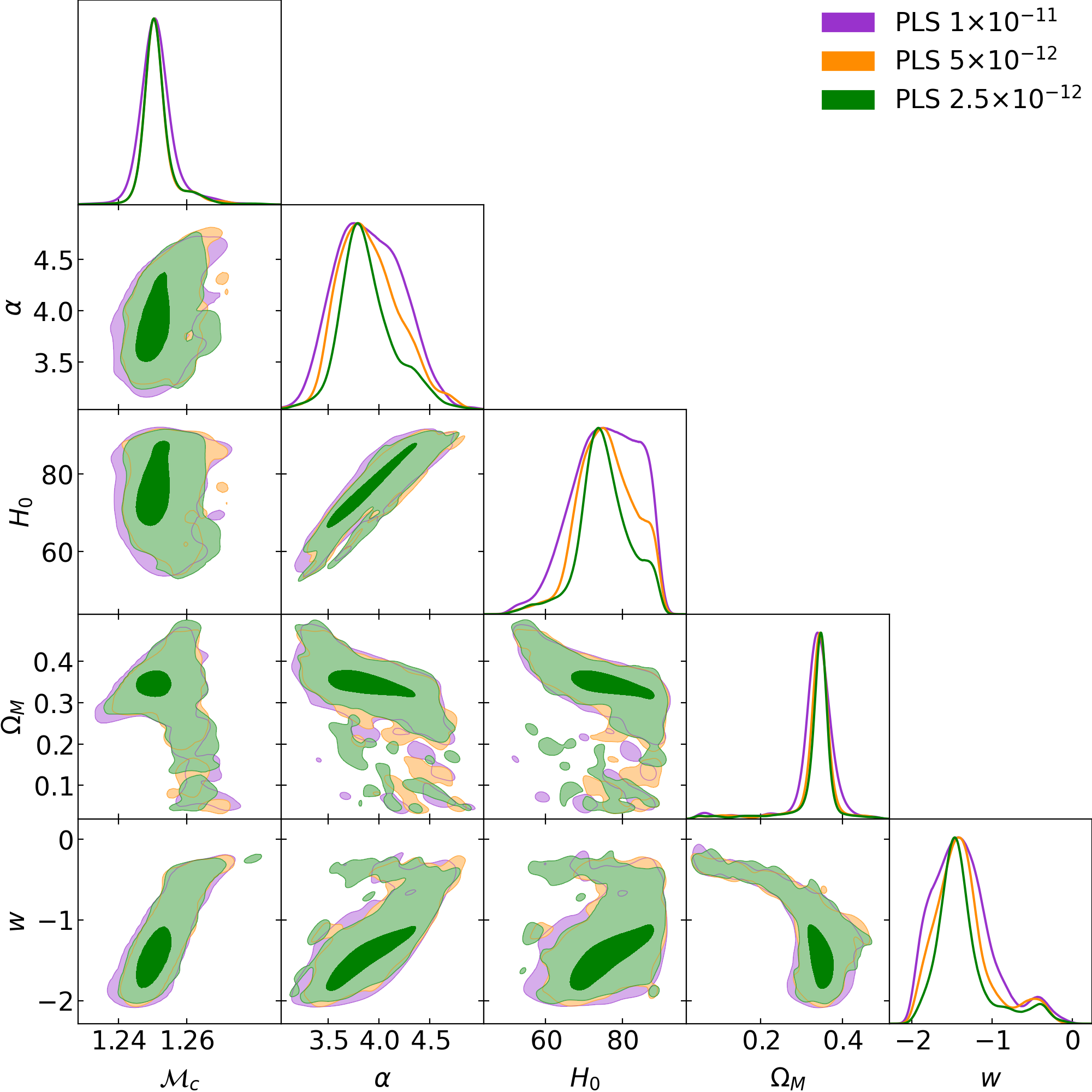}
\caption{\label{fig:posteriors_cosmo_tot} Joint constraints (68$\%$ and 95$\%$ confidence regions) and marginalized posterior distributions on all the considered parameters, $\mc$, $\alpha$,  $H_{0}$, $\Omega_{M}$ and $w$, for three different sensitivity levels in the kHz range. The input values for the parameters are $\{ 1.25 \, M_{\odot}, 3.8, 73\, \rm{km} \, \rm{s}^{-1} \rm{Mpc}^{-1}, 0.315, -1.5\}$. }
\end{figure}

We also investigate the constraining power of our mock data-set when incorporating the other cosmological parameters, $\Omega_{M}$ and $w$. In Fig.~\ref{fig:posteriors_cosmo_tot}, we show the joint constraints (68$\%$ and 95$\%$ confidence regions) and marginalized posterior distributions on our full set of parameters $\astro$ and $\cosmo$. We also report the associated marginalized percentage constraints at the 68$\%$ confidence level in Table~\ref{tab:perc_constraints_5}.
\begin{table}[b]
\caption{\label{tab:perc_constraints_5}%
Marginalized percentage constraints at the 68$\%$ confidence level on all the considered parameters, with input values $\{ 1.25 \, M_{\odot}, 3.8, 73\, \rm{km} \, \rm{s}^{-1} \rm{Mpc}^{-1}, 0.315, -1.5\}$, for the three different kHz sensitivity levels.}
\begin{ruledtabular}
\begin{tabular}{cccc}
 & PLS $1\times 10^{-11}$ & PLS $5\times 10^{-12}$ & PLS $2.5\times 10^{-12}$\\
\colrule 
$\mc$    &   $0.4\%$  &  $0.3\%$   &   $0.3\%$  \\ [3pt]
$\alpha$ &  $8.7\%$   &   $7.5\%$  &  $6.2\%$   \\ [3pt]
$H_{0}$  &  $11\%$   &  $9,7\%$   &  $8.2\%$   \\ [3pt]
$\Omega_{M}$  &  $9.4\%$   &  $7.1\%$   &  $6.5\%$   \\ [3pt]
$w$  &  $25\%$   &  $21\%$   &  $17\%$   \\ [3pt]
\end{tabular}
\end{ruledtabular}
\end{table}
As expected, including a larger number of parameters in our model leads to broader posterior constraints. The increased complexity of the parameter space results in multimodal posterior distributions with several secondary peaks and introduces a higher level of degeneracy, especially for the parameters $\alpha$ and $H_{0}$. Nevertheless, the constraining power of our data-set remains significant, as the posteriors add information with respect to the priors for all parameters and at all kHz sensitivity levels. Specifically, the primary peak of the marginalized posterior constraint on $H_{0}$, despite its asymmetry, has a FWHM of approximately $10$ for PLS $ = 2.5 \times 10^{-12}$. Therefore, such a measurement could potentially exclude one of the existing estimates at a $68\%$ confidence level.

\section{\label{sec:discussion}Discussion and conclusions}

In this paper, we delve into the insights that the high-frequency features of the SGWB produced by coalescing BNSs can offer in understanding the underlying astrophysics and cosmology. Specifically, we studied the constraining capabilities of mock SGWB measurements in the kHz regime. In the high-frequency range, the SGWB energy density shows a distinctive peak that contains most of the physical information. There are several stellar, galactic, and cosmological processes that affect the amplitude and shape of the SGWB. However, within the frequency range explored by ground-based interferometers, the SGWB from merging BNSs follows a power-law behavior with a fixed $f^{2/3}$ slope. As a result, SGWB measurements in this region only allow for the determination of the signal's amplitude, leading to considerable degeneracy among the physical factors responsible for its production. In contrast, the frequency band above a few hundred Hz offers a unique opportunity to probe the distinct peak of the SGWB, allowing us to constrain the astrophysical and cosmological processes that generate the signal.

As a first step, we identified a set of astrophysical and cosmological parameters that effectively characterize the SGWB sources. We focused our analysis on the SGWB generated by coalescing BNSs, instead of BBHs and NSBHs, because they are minimally affected by metallicity and mainly depend the redshift evolution of the galaxy main sequence, which is well constrained. We adopt empirical, data-driven prescriptions for the galactic environment and restrict our selection of astrophysical parameters to the stellar domain. We use only two astrophysical parameters to describe the BNS population, $\astro = \{ \mc, \alpha \}$, where $\mc$ is the chirp mass at which the BNS mass distribution peaks, and $\alpha$ is the common envelope efficiency parameter. 
On the cosmological side, both amplitude and shape of the SGWB depend on the adopted scenario. Each cosmology is defined by specific parameters, either directly or indirectly influencing the expression for the energy density of the SGWB, as given in Eq.~(\ref{eq:omega_gw}). Specifically, we work with the parameters $\cosmo = \{H_{0}, \Omega_{M}, w\}$, where $H_{0}$ is the Hubble parameter, $\Omega_{M}$ the matter density parameter, and $w$ the dark energy equation of state parameter.
We first investigated how varying these parameters affects the SGWB energy density. Notably, we found that both astrophysical and cosmological factors have similar order-of-magnitude effects on the SGWB energy density. Then, we performed an MCMC analysis using a set of mock data covering a frequency range between a few tens of Hz and a few kHz. The main goal was to evaluate the constraining power of these data on our set of astrophysical and cosmological parameters. For the data points in the $\sim 10$ Hz range, we set the errors to match the PLS of ET. In the kHz range, instead, we assumed progressively lower errors to investigate the minimum sensitivity requirements in this frequency band that would enable effective constraints on all the selected astrophysical and cosmological parameters. 
Restricting the analysis only to the parameters $\{ \mc, \alpha, H_{0}\}$, we discovered that our mock data had constraining power for PLSs lower than $10^{-11}$ in the kHz frequency band. With a PLS of $5\times 10^{-12}$ and $2.5\times 10^{-12}$, we could retrieve the Hubble parameter with a precision that has the potential to solve the Hubble tension at 1$\sigma$ and 2$\sigma$, respectively.
Including also the remaining parameters, $\Omega_{M}$ and $w$, we observed a decay in the constraining power. The increased complexity of the parameter space leads to the emergence of several secondary peaks in the posterior distributions. Despite this, the data still add valuable information to the priors, offering potential insights into the values of our astrophysical and cosmological parameters.

In conclusion, our science case establishes the relevance of the SGWB generated by BNSs as a robust observational tool within the kHz frequency range. Its characteristic peak contains a significant amount of physical information, enabling effective constraints on many astrophysical and cosmological processes involved in the production of the SGWB. Despite the complex interplay among numerous parameters, this observable remains effective in providing valuable insights, when measured with sufficient precision.

\begin{acknowledgments}
We acknowledge valuable input from the anonymous referee. We warmly thank Michele Maggiore for carefully reading the manuscript and for useful discussions, Enis Belgacem for providing us with the official ET sensitivity curves, and Lumen Boco and Carole Périgois for their helpful feedback. GC and CB acknowledge partial support by the INDARK INFN grant. AL acknowledges funding from: the EU H2020-MSCA-ITN-2019 Project 860744 \textit{BiD4BESt: Big Data applications for black hole Evolution STudies}, PRIN MUR 2022 project n. 20224JR28W "Charting unexplored avenues in Dark Matter"; INAF Large Grant 2022 funding scheme with the project "MeerKAT and LOFAR Team up: a Unique Radio Window on Galaxy/AGN co-Evolution; INAF GO-GTO Normal 2023 funding scheme with the project "Serendipitous H-ATLAS-fields Observations of Radio Extragalactic Sources (SHORES)". CB acknowledges support from the COSMOS $\&$ LiteBIRD Networks by the Italian Space Agency (\url{http://cosmosnet.it}). AL and MS are further supported by: ``Data Science methods for MultiMessenger Astrophysics \& Multi-Survey Cosmology'' funded by the Italian Ministry of University and Research, Programmazione triennale 2021/2023 (DM n.2503 dd. 9 December 2019), Programma Congiunto Scuole; Italian Research Center on High Performance Computing Big Data and Quantum Computing (ICSC), project funded by European Union - NextGenerationEU - and National
Recovery and Resilience Plan (NRRP) - Mission 4 Component 2 within the activities of Spoke 3 (Astrophysics and Cosmos Observations).
\end{acknowledgments}

\bibliography{bibliography}
\bibliographystyle{apsrev4-1}

\end{document}